
\magnification=1200
\vsize=25truecm
\hsize=16truecm
\baselineskip=0.6truecm
\parindent=1truecm
\nopagenumbers
\font\scap=cmcsc10
\hfuzz=0.8truecm
\def\hdot{\!\cdot\!}

\null \bigskip
\centerline{\bf MULTILINEAR OPERATORS: THE NATURAL EXTENSION}
\centerline{\bf OF HIROTA'S BILINEAR FORMALISM}

\vskip 2truecm
\centerline{\scap B. Grammaticos}
\centerline{\sl LPN, Universit\'e Paris VII}
\centerline{\sl Tour 24-14, 5${}^{\grave eme}$ \'etage}
\centerline{\sl 75251 Paris, France}
\vskip 1truecm
\centerline{{\scap A. Ramani} and {\scap J. Hietarinta $\! ^*$}}
\centerline{\sl CPT, Ecole Polytechnique}
\centerline{\sl 91128 Palaiseau, France}
\vskip 5truecm
\bigskip
\bigskip
\noindent Abstract
\smallskip
\noindent We introduce multilinear operators, that generalize Hirota's
bilinear $D$ operator, based on the principle of gauge invariance of
the $\tau$ functions. We show that these operators can be constructed
systematically using the bilinear $D$'s as building blocks. We
concentrate in particular on the trilinear case and study the possible
integrability of equations with one dependent variable. The 5th order
equation of the Lax-hierarchy as well as Satsuma's lowest-order gauge
invariant equation are shown to have simple trilinear expressions. The
formalism can be extended to an arbitrary degree of multilinearity.

\vfill
\hrule\smallskip
\noindent$^*$ On leave of absence from Department of Physics, University
of Turku, FIN-20500
Turku, Finland
\bigskip\eject

\footline={\hfill\folio}
\pageno=2

{\scap 1. Introduction}
\medskip
\noindent The Hirota bilinear operators were introduced as an
antisymmetric extension to the usual derivative [1], because of their
usefulness for the computation of multisoliton solution of nonlinear
evolution equations. The bilinear operator
$D_x\equiv\partial_{x_1}-\partial_{x_2}$ acts on a pair of functions
(the `dot product') antisymmetrically: $$D_xf\hdot
g=(\partial_{x_1}-\partial_{x_2})f(x_1)g(x_2)\big
|_{x_2=x_1=x}=f'g-g'f.\eqno(1)$$ The Hirota bilinear formalism has
been instrumental in the derivation of the multisoliton solutions of
(integrable) nonlinear equations. A prerequisite to its application is
a dependent variable transformation that converts the nonlinear
equation into a quadratic ``prepotential'' form. This is best
understood in a specific example, so let us consider the paradigmatic
case of the KdV equation. Starting from
$$
u_{xxx}+6uu_x+u_t=0,\eqno(2)
$$
we introduce the transformation $u=2\partial^2_x\log F$ and obtain
(after one integration):
$$
F_{xxxx}F-4F_{xxx}F_{x}+3F^2_{xx}+F_{xt}F-F_xF_t=0.\eqno(3)
$$
This last equation can be written in a particularly condensed form
using the Hirota $D$ operator:
$$
(D^4_x+D_xD_t)F\hdot F=0.\eqno(4)
$$
The power of the bilinear formalism lies in the fact that for
multi-soliton solutions the $F$'s are simple polynomials of
exponentials [2]. Thus the construction of soliton solutions becomes
an algebraic problem. This approach has made possible the
investigation of large classes of bilinear equations and the
classification of integrable cases [3]. (The integrability of these
equations has been confirmed by singularity analysis [4]).
\bigskip

{\scap 2. Gauge-invariant bilinear operators}
\medskip
\noindent
An important observation (that has motivated the present line of
research) is the relation of the ``physical'' variable
$u=2\partial^2_x\log F$ (for the KdV equation) to the Hirota's
function $F$: the gauge transformation $F\to e^{px+\omega t} F$ leaves
$u$ invariant.  It turns out that this is a general property of
bilinear equations. In fact, one can {\sl define} the Hirota's
bilinear equations through the requirement of gauge invariance.  We
will now prove this statement.

Let us introduce a general bilinear expression and ask that it be
invariant under the gauge transformation $F\to e^{\eta} F$ with $\eta
=px+\omega t$:
$$
\sum^N_{k=0} c_k(\partial^ke^\eta
f)(\partial^{N-k}e^\eta g)=e^{2\eta}\sum^N_{k=0} c_k(\partial^k
f)(\partial^{N-k} g).\eqno(5)
$$
Expanding the left hand side and equating the coefficients of
$(\partial^k f)(\partial^{N-k} g)$ we obtain, for all $n,m$:
$$
\sum^N_{k=0} \pmatrix{k\cr n\cr}
\pmatrix{N-k\cr m\cr} p^{N-n-m}c_k=c_n
\delta_{N-n-m},\eqno(6)
$$
where $\pmatrix{k\cr n\cr}$ is the binomial coefficient.  From the
structure of the left hand side, we have the inequalities $0\le n\le
k\le N-m$.  Taking $m=N-n$ we find that $k=n$ and (6) is identically
satisfied. For all other values of $m$ ($m=$0, 1, \dots, $N-n-1$) we
must find those $c_k$'s that satisfy
$$
\sum^N_{k=0} \pmatrix{k\cr
n\cr} \pmatrix{N-k\cr m\cr} c_k=0.\eqno(7)
$$
In the special case $m=N-n-1$ (7) reduces to:
$$(N-n)c_n+(n+1)c_{n+1}=0,$$
the solution of which is
$$c_n=(-1)^n\pmatrix{N\cr n\cr}c_0.\eqno(8)$$
This solution does, in fact, satisfy (7) for {\sl all} values of $m$:
using (8) we find that the lhs of (7) is just the expression of
$(1-1)^{N-n-m}c_0$ and thus equal to zero. So the {\sl only} gauge
invariant bilinear differential operator is (up to a scaling of $c_0$)
$\sum^N_{k=0} (-1)^k \pmatrix{N\cr k\cr} \partial_1^k\partial_2^{N-k}
=(\partial_1-\partial_2)^N$, i.e. the Hirota operator $D^N_{12}$. (The
use of the indices (1,2) may appear superfluous at this stage, since
in the bilinear case there are only two variables on which $D$
operators act. However the notation will be particularly useful in the
higher multilinear cases).
\bigskip

{\scap 3. Gauge-invariant trilinear operators}
\medskip
\noindent
The Hirota bilinear operators can be obtained on the sole requirement
(admittedly a strong one) of gauge invariance. In this paper our
objective is the extension of the bilinear formalism and the
introduction of multilinear operators. Very few results exist in this
direction.  Satsuma and collaborators have introduced a particular
class of trilinear equations, that can be written as a single
$(3\times 3)$ determinant [5]. A full hierarchy of equations was
obtained and the richness of the solutions presented are a strong
indication of their integrability. However, Satsuma's approach offers
no clue on what a trilinear {\sl operator} should be. Here we will use
the same principle used to find the bilinear $D$'s: gauge invariance.

For trilinear expressions the invariance condition writes:
$$
\sum_{k+l+m=N} c_{klm}(\partial^ke^\eta f)(\partial^le^\eta g)
(\partial^me^\eta h)=e^{3\eta}\sum_{k+l+m=N}
c_{klm}(\partial^k f)(\partial^l g)(\partial^m h).\eqno(9)
$$
In analogy to the bilinear case we find:
$$\sum_{k+l+m=N} c_{klm}\pmatrix{k\cr \kappa\cr}\pmatrix{l\cr
\lambda\cr}\pmatrix{m\cr
\mu\cr}p^{k+l+m-\kappa-\lambda-\mu} =c_{\kappa \lambda \mu}
\delta_{\kappa+\lambda+\mu-N}.
\eqno(10)
$$
For  $\kappa+\lambda+\mu=N$ equation (10) is identically satisfied
and we are left with
$$
\sum_{k+l+m=N} c_{klm}\pmatrix{k\cr \kappa\cr}\pmatrix{l\cr
\lambda\cr}\pmatrix{m\cr
\mu\cr} =0 \quad (\rm{with}\quad \kappa+\lambda+\mu < N).\eqno(11)
$$
Consider the $\kappa+\lambda+\mu =N-1$ equations. They write:
$$
(\mu+1)c_{\kappa\lambda\mu+1}+(\kappa+1)c_{\kappa+1\lambda\mu}+
(\lambda+1)c_{\kappa\lambda+1\mu}=0.\eqno(12)
$$
We have $N(N+1)\over 2$ such equations for the $(N+1)(N+2)\over 2$,
$c$'s.  Thus these equations would determine the $c$'s up to $N+1$
free coefficients, provided the rank of the system is maximal. This is
indeed the case. In fact, the equations ($\kappa,\lambda,\mu=0$)
contain for the first time $c_{\kappa\lambda1}$ (which can be
expressed in terms of the $(N+1)$ $c_{\kappa\lambda0}$). We can then
compute successively the higher-$\mu$ terms up to the last equation
($\kappa=0,\lambda=0,\mu=N-1$) from which we can solve for
$c_{00N-1}$. Thus all $c_{\kappa\lambda\mu}$ for $\mu>0$ can be
expressed in terms of the $c_{\kappa\lambda 0}$.  So, given that the
rank is maximal, we can choose {\sl any} basis for the $c$'s. A most
convenient basis are the following $N+1$ operators:
$(\partial_1-\partial_2)^n(\partial_1-\partial_3)^{N-n}$ for
$n=0,\dots,N$.

Thus the basic building blocks for the trilinear operators are again
the Hirota bilinear $D$'s, we must just specify the indices in this
case. We thus have $D_{12}\equiv\partial_{x_1} -\partial_{x_2}$,
$D_{23}\equiv\partial_{x_2} -\partial_{x_3}$,
$D_{31}\equiv\partial_{x_3} -\partial_{x_1}$, but, of course the three
are not linearly independent: $D_{12}+D_{23}+D_{31}=0$.  Their action
on a `triple dot product' is analogous to the bilinear case:
$$
D_{12}f\hdot g\hdot
h=(\partial_{x_1}-\partial_{x_2})f(x_1)g(x_2)h(x_3)\big
|_{x_3=x_2=x_1=x}=(f'g-fg')h.\eqno(13)
$$
The choice of a particular pair of $D$'s as the basic trilinear
operators breaks the symmetry between the three coordinates
$x_i$'s. It is possible to restore this symmetry by introducing a
different basis for the trilinear operators, $T$ and $T^*$:
$$
T=\partial_1+j\partial_2+j^2\partial_3\, , \quad
T^*=\partial_1+j^2\partial_2+j\partial_3,
\eqno(14)
$$
where $j$ is the cubic root of unity, $j=e^{2i\pi/3}$. (Note that the
star in $T^*$ indicates complex conjugation for the coefficients in
$T$ but not for the independent variables). The price we have to pay
for restoring this symmetry is that the operators are now more
complicated. Note that $T^nT^{*m}F\hdot F\hdot F=0$ unless $n-m\equiv
0$ (mod 3), which is the equivalent to the bilinear property $D^n
F\hdot F=0$ unless $n\equiv 0$ (mod 2).

The generalization to higher multilinear equations is
straightforward. One can introduce the set of $n(n-1)/2$ operators
$D_{ij}$ acting on $n$-tuple dot-products $D_{ij}f_1\hdot f_2\!\cdot
\dots \cdot\! f_n.$ (Of course only $n-1$ of the $D_{ij}$'s are
independent, a convenient basis being the $D_{1j},$ $j=2, \dots
n)$. As in the trilinear case, one can also construct ``symmetric''
operators:
$$M_m=\sum_{k=1}^{n-1} z_m^k \partial_k\eqno(15)$$
where
the $z_m$'s are the $(n-1)$ $n$-th roots of unity other than one.
\bigskip

{\scap 4. Examples of multilinear equations}
\medskip
\noindent
Multilinear operators are not just a trivial extension of the Hirota
bilinear formalism. They are necessary for the description of
nonlinear evolution equations that cannot be cast in a bilinear form
and such equations do exist. An interesting example is the fifth-order
equation of the Lax hierarchy [6]:
$$
u_{xxxxx}+10uu_{xxx}+20u_{x}u_{xx}+30u^2u_x+u_t=0.\eqno(16)
$$
While this Lax-5 equation does not possess a simple bilinear
expression like KdV itself, it has a trilinear form (with
$u=2\partial^2_x\log F$):
$$
(7T_x^6+20T_x^3T_x^{*3}+27T_xT_t)F\!\hdot F\!\hdot F=0.\eqno(17)
$$
In the previous section we referred to
Satsuma's trilinear equations [5]. The lowest-order one
$$
\left|\matrix{ F_{yy}&F_y&F_{xy}\cr F_y&F&F_x\cr
F_{xy}&F_x&F_{xx}\cr }\right|=0,\eqno(18)
$$
(equivalent, through
$F=e^w$, to the Monge-Amp\`ere equation $w_{xy}^2-w_{xx}w_{yy}=0$) is
gauge-invariant and can be written as:
$$
(T_xT_x^*T_yT_y^*-T_x^2T_y^{*2})F\!\hdot F\!\hdot F=0.\eqno(19)
$$
(The higher Satsuma equations are given by sets of equations with
``dummy'' independent variables and it is not clear how to implement
the gauge-invariance requirement in such a situation).

The equation:
$$
\displaylines{F^2F_{xxxxy}-FF_yF_{xxxx}-4FF_xF_{xxxy}+2FF_{xx}F_{xxy}+
4F_xF_yF_{xxx}-2F_yF_{xx}^2\hfill\cr\hfill-4F_xF_{xx}F_{xy}+4F_x^2F_{xxy}
+4(F^2F_{xxt}-FF_{xx}F_t-2FF_xF_{xt}-2F_x^2F_t)=0,\quad(20)\cr}
$$
obtained as a reduction of a self-dual Yang-Mills equations [7], can
also be written as a genuinely trilinear equation:
$$
(T_x^4T_y^{*}+8T_x^3T_x^{*}T_y-36T_x^2T_t)F\!\hdot F\!\hdot F=0.\eqno(21)
$$

Further examples can be presented and we can, of course, construct
also higher multilinear equations (quadri-, penta-, etc.)
equations. Instead of dealing with specific cases let us present here
some general considerations.  Let us start with a nonlinear (in $u$)
equation, of order $k+1$ having the form $u_t+\partial_x
P(u,u_x,\dots,u_{kx})=0$. Several well known integrable equations
belong to this class. It is simpler to work with the time independent
part $P(u,u_x,\dots,u_{kx})=0$ which only involves derivatives up to
order $k$.  We consider the leading part of $P$ which we assume to be
weight-homogeneous in $u$ and $\partial_x$, with $u$ having the same
weight as $\partial_x^2$. Then we can transform this leading part to a
multilinear expression through the transformation $u=\alpha
\partial_x^2(\log F)$ and obtain generically an $k+2$-multilinear
equation. The scaling factor $\alpha$ can then be chosen so as to make
the lowest-derivative terms vanish, that is those that appear under
the combination $(F'^{k+2}-{k+2\over 2}F''F'^{k}F)$. (This is possible
because these two terms have a common factor, polynomial in
$\alpha$). Then an $F^2$ term can be factored out and the resulting
multilinear equation is at most $k$-linear.

At order five ($k=4$) we have three integrable equations. We should
expect, in principle, these equations to have quadrilinear forms. Some
unexpected cancellations, however, do occur. Thus the Sawada-Kotera
[8] equation has a bilinear expression, the Lax-5 has the trilinear
form we gave in (17), but the Kaup-Kuperschmidt [9] is
quadrilinear. At order seven ($k=6$) three integrable equations were
known to exist. The higher Sawada-Kotera has a trilinear expression
(see next section) while the 7th order equation in the Lax hierarchy
is a pentalinear one. Again, for the higher Kaup-Kuperschmidt equation
no extra simplification is possible and this equation is hexalinear.
For all these equations the time derivative can be incorporated in the
multilinear equations in a very simple way without altering the degree
of multilinearity.

Thus the multilinear extension to Hirota bilinear approach has a wide
range of applicability (in particular if we allow for multicomponent
equations, introducing more than one dependent functions, in analogy
to the bilinear case).

\bigskip
{\scap 5. Singularity analysis of trilinear equations}
\medskip
\noindent
All the above equations have as a common characteristic their
integrability. The study of integrability is, in fact, the motivation
behind the multilinear approach. The systematic classification of
bilinear equations we presented in [3,4] was based on the study of
multisoliton solutions and of the Painlev\'e property. We intend to
come back to the investigation of soliton solutions for our
multilinear equations in some future work. In the following paragraph
we will limit ourselves to the singularity analysis of trilinear
equations involving only one dependent variable, i.e. unicomponent
equations. (Let us recall here that in the bilinear case the study of
these simplest equations led us to conclusive results).

In order to perform the Painlev\'e analysis, we shall study the
leading (highest-order) part of the equations with just one
independent variable. This is sufficient for the computation of
dominant singularities and resonances, although for the check of
resonance compatibility we would need the full equation, which remains
unspecified at this stage. (For example, for the analysis of an
equation like (17), we would consider only the $T^6,T^3T^{*3}$ terms
and not $T_xT_t$, since the influence of the latter would appear only
at the resonance condition).

Since the dependent function $u$ in a nonlinear equation is related to
the multilinear $F$ through $u=2\partial^2\log F$ it is clear that a
zero in $F$ induces a pole-like behaviour in $u$.  Let us show how one
performs the singularity analysis for trilinear equations in a
concrete example: $T^2T^{*2}f\hdot f\hdot f=0$. Putting $f\sim x^n$
(for the dominant part) we find $n=0,1$ as the only possible
behaviours. The first corresponds to a nonsingular Taylor-like
expansion, which is always possible. The second behaviour $f\sim x$
corresponds to a simple zero that would give a (double) pole in
$u$. The resonances in this case are obtained if we substitute
$f=x(1+\phi x^r)$ and collect terms linear in $\phi$.  The result is
$r=-1,0,1,6$ and (after a check that no incompatibilities arise at any
resonance) we conclude that this equation passes the Painlev\'e
test. This is what one would have expected, had we looked at the
nonlinear form of the equation, $u_{xx}+3u^2=0$, which is just the
time-independent part of KdV integrated once.

We shall not present the details of the singularity analysis of all
the equations that we have studied. The results are summarized
below. The notation we are using is the following: $E(n,m)$, (which is
identical to $E(m,n)$), represents the expression $T^mT^{*n}F\hdot
F\hdot F$.  Note that for a given $N=n+m$ there may exist several
pairs of $(n,m)$ such that $E(n,m)$ is not identically zero, namely
those for which $n\equiv m$ (mod 3). The leading part of the general
equation at order $N$ is then given by a linear combination of all the
non-vanishing $E(n,m)$'s. In each case we give below the precise
combinations that lead to equations with the Painlev\'e property. The
nonlinear forms of the equations are obtained by the standard
substitution $F=e^g$ followed by $u=2g''$.

\noindent $N=2: \quad E(1,1)$
\item{} In this case we can write the result also in bilinear form:
\item{} $E(1,1)\propto F (D^2 F\hdot F)=2F(F''F-F'^2)=2e^{3g} g''=e^{3g} u$.

\noindent $N=3: \quad E(3,0)$
\item{} $E(3,0)\propto F'''F^2-3F''F'F+2F'^3 = e^{3g} g'''\propto e^{3g} u'$.

\noindent $N=4: \quad E(2,2)$
\item{} Here also we can write the result in bilinear form:
\item{} $E(2,2)\propto F (D^4 F\hdot F) = 2e^{3g} (g''''+6g''^2)=e^{3g}
(u''+3u^2)$.
\item{} This, of course, is just the leading part of the KdV equation
in potential form.

\noindent $N=5: \quad E(4,1)$
\item{} $E(4,1)/F^3\propto u'''+6u'u$.
\item{} Note that this is the derivative of the expression obtained at $N=4$.

\noindent $N=6: \quad \lambda E(6,0) +\mu E(3,3)$. This is the first case
where we have two
possible $n,m$ pairs. The $\lambda,\mu$ combinations that pass the
Painlev\'e test are the
following
\item{a)} $(7E(6,0) +20E(3,3))/F^3\propto u''''+10u''u+5u'^2+10u^3$.
\item{} This is the leading part of the 5th order equation in the Lax
hierarchy of KdV, eq.(16),
integrated once.
\item{b)} $(-2E(6,0) +20E(3,3))/F^3\propto u''''+15u''u+15u^3$.
\item{} This is the leading part of the Sawada-Kotera equation,
integrated once.
\item{c)} $(E(6,0) -E(3,3))/F^3\propto uu''-u'^2+u^3$.
\item{} The trilinear form of this case is $F''''(F''F-F'^2)-F'''^2F
+2F'''F''F-F''^3$, which can
be cast in determinantal form and is a 1-dimensional member of the
Satsuma family:
$$\left|\matrix{ F''''&F'''&F''\cr F'''&F''&F'\cr F''&F'&F\cr }\right|$$

\noindent $N=7: \quad E(5,2)$
\item{} $E(5,2)/F^3\propto u^{(5)}+15u'''u+15u''u'+45u'u^2$.
\item{} This is the Sawada-Kotera equation, i.e. the derivative of the
expression obtained
in $N=6$b.

\noindent $N=8: \quad \lambda E(7,1) +\mu E(4,4)$ and we take $u=6g''$
instead of $u=2g''$
used before
\item{a)} $(4E(7,1) +5E(4,4))/F^3\propto
u^{(6)}+6u''''u+10u'''u'+5u''^2+10u''u^2+10u'^2u+{5\over 3}u^4$
\item{} This would correspond to a $7^{th}$ order equation which, we
believe, leads to a new
integrable case.
\item{b)} $(4E(7,1) +14E(4,4))/F^3\propto
u^{(6)}+7u''''u+7u'''u'+7u''^2+14u''u^2+7u'^2u+{7\over 3}u^4$
\item{} This is the higher Sawada-Kotera equation we referred to in
the previous section.
\item{c)} $(E(7,1) -E(4,4))/F^3\propto u''''u-3u'''u'+2u''^2+4u''u^2-
3u'^2u+{2\over3}u^4$
\item{} This new equation looks like an extension of Satsuma's equation
given at $N$=6c above,
but
it cannot be written as a single determinant.

\noindent At order $N=9$ there are no cases passing the Painlev\'e test.

\noindent $N=10: \quad \lambda E(8,2) +\mu E(5,5)$ and we take $u=30g''$
\item{} $(5E(8,2) +4E(5,5))/F^3\propto
u^{(8)}+2u^{(6)}u+4u^{(5)}u'+6u''''u''+5u'''^2+{6\over 5}u''''u^2+
4u'''u'u+2u''^2u+2u''u'^2
+{4\over 15}u''u^3+{2\over 5}u''^2u^2+{2\over 375}u^5$
\item{} This is also a new equation.

\noindent Furthermore, no integrable candidates were found at orders
$N=11,12$. The singularity analysis at these higher orders becomes
progressively more difficult.  (Already at $N=12$ there exist three
nonvanishing $n,m$ combinations). It is, thus, not possible to extend
our investigation to very high orders (as was done in our study of
bilinear equations [4]), but we do believe that no further integrable
candidates exist at higher orders. As in the bilinear case, a finite
(and rather small) number of unicomponent trilinear equations possess
the Painlev\'e property and can thus be integrable.
\bigskip

{\scap 6. Conclusion}
\medskip
\noindent
In the preceding paragraphs we have presented an extension of Hirota's
bilinear formalism that can encompass any degree of
multilinearity. The main guide in our investigation has been the
requirement that the equations be gauge-invariant. Since our primary
objective is the study of integrability, we have also presented a
classification of one-component trilinear equations that pass the
Painlev\'e test. The crucial difference between the bilinear and the
tri- (and multi-)linear case(s) is that now free parameters enter
already at the leading part. This means that the Painlev\'e analysis
of the higher order unicomponent equations becomes increasingly
difficult.  Once the leading parts of these equations are fixed, one
can study the lower-order terms that can be added without destroying
the Painlev\'e property.  Starting from a complete classification of
unicomponent equations one can build up multicomponent ones following
the approach we presented in [3] for the bilinear case.  Another
interesting direction would be the computation of the multisoliton
solutions of the trilinear equations. This would furnish another check
for their possible integrability.  Clearly, the domain of multilinear
equations is still a {\sl terra incognita} that deserves serious
study.

\bigskip
{\scap Acknowledgements}

\noindent  The authors are grateful to R. Hirota, J. Satsuma and
W. Oevel for illuminating
discussions and
 exchange of correspondence.
\bigskip
{\scap References}
\medskip
\item{[1]} R. Hirota, Phys. Rev. Lett. 27 (1971) 1192.
\item{[2]} R. Hirota in ``{\it Soliton}'' eds.  R. K. Bullough and
P. J. Caudrey, p 157, Springer
(1980)
\item{[3]} J. Hietarinta, J. Math. Phys. 28 (1987) 1732, 2094, 2586;
ibid 29 (1988) 628.
\item{[4]} B. Grammaticos,  A. Ramani and J. Hietarinta , J. Math.
Phys. 31 (1990) 2572
\item{[5]} J. Matsukidaira, J. Satsuma and W. Strampp, Phys. Lett.
A 147 (1990) 467.
\item{[6]} H.C. Morris, J. Math. Phys. 18 (1977) 530.
\item{[7]} J. Schiff in ``{\it Painlev\'e Transcendents, Their
Asymptotics and Physical
Applications}'', eds. D. Levi and P. Winternitz, p 393 Plenum (1992).
\item{[8]} K. Sawada and T. Kotera, Progr. Theor. Phys. 51 (1974) 1355.
\item{[9]} D.J. Kaup, Studies Appl. Math. 62 (1980) 189.
\item{} B.A. Kupershmidt and G. Wilson, Invent. Math. 62 (1981) 403.
\end